# Molecular structural order and anomalies in liquid silica

by

M. Scott Shell, Pablo G. Debenedetti[*], and Athanassios Z. Panagiotopoulos

Department of Chemical Engineering, Princeton University, Princeton, NJ 08544

**Abstract**

The present investigation examines the relationship between structural order, diffusivity anomalies, and density anomalies in liquid silica by means of molecular dynamics simulations. We use previously defined orientational and translational order parameters to quantify local structural order in atomic configurations. Extensive simulations are performed at different state points to measure structural order, diffusivity, and thermodynamic properties. It is found that silica shares many trends recently reported for water [J. R. Errington and P. G. Debenedetti, Nature **409**, 318 (2001)]. At intermediate densities, the distribution of local orientational order is bimodal. At fixed temperature, order parameter extrema occur upon compression: a maximum in orientational order followed by a minimum in translational order. Unlike water, however, silica's translational order parameter minimum is broad, and there is no range of thermodynamic conditions where both parameters are strictly coupled. Furthermore, the temperature-density regime where both structural order parameters decrease upon isothermal compression (the structurally anomalous regime) does not encompass the region of diffusivity anomalies, as was the case for water.

---

[*] Corresponding author; email: pdebene@princeton.edu

I.  **Introduction**

Despite their ubiquity in everyday life, network-forming fluids and their anomalous behavior continue to pose challenging research questions. Network-forming fluids exhibit strong, orientation-dependent intermolecular interactions. In the liquid phase, these interactions can promote the formation of open, locally ordered environments which are responsible for unusual behavior not observed in simple liquids. Two "canonical" examples of network-forming fluids are water and silica. That liquid water is vitally important in virtually all aspects of our lives is uncontested. Amorphous silica, formed by rapid cooling of the liquid melt, is also fundamental: over 60% of the earth's crust is made of it, and this material is responsible for numerous optical technologies and electrical devices, as well as being used as an additive in paints, soaps, and foods [1].

Due to their network-forming ability, the liquid phases of water and silica exhibit a number of peculiar properties which have been observed in experiment and simulation. Most familiar is water's well-known density maximum at 4 °C, below which temperature the liquid exhibits a negative thermal expansion. Silica shows similar behavior [2, 3]. Associated anomalies are isobaric heat capacity minima and isothermal compressibility minima, experimentally observed in water [4, 5] and seen in computer simulated silica [6-8]. Anomalous behavior is also observed in kinetic phenomena; in both liquids at sufficiently low temperatures, a range of pressure exists for which diffusivity increases and viscosity decreases upon compression [9-11]. Both water and silica have two distinct amorphous states, denoted high-density amorphous (HDA) and low-density amorphous (LDA). HDA and LDA states have been observed experimentally in water [12] and through simulation in silica [13-15]. Direct experimental evidence of the transition in silica has only recently been reported [16]. Although the nature of this transition has not been firmly established, the preponderance of evidence suggests that it is a first-order transition, i.e., that it entails discontinuities in density and enthalpy [17]. Recent work also indicates that water and silica exhibit a "fragile-to-strong" transition upon cooling [8, 18]. That is, the behavior of molecular relaxation times is Arrhenius at low temperatures (strong behavior) and at sufficiently high temperature "super-Arrhenius," i.e., positive curvature occurs when the log of the relaxation time is plotted as a function of inverse temperature (fragile behavior.)

Because the locally ordered environments of network-forming fluids are thought to play a role in imparting their unusual properties, the relationship between molecular structural order and thermophysical property anomalies is of great interest. The introduction of simple geometric measures that quantify microscopic structural order holds promise for clarifying this connection. When considering only molecular centers, two such order parameters are useful: a translational order parameter measuring the tendency of pairs of molecules to be separated by preferential distances, and an orientational order parameter measuring the tendency of a central molecule and its nearest neighbors to adopt preferential mutual orientations (e.g., the tetrahedral angle.) The average values of these parameters are functions of state, and the mean translational and orientational order as a function of density and temperature can be evaluated alongside thermodynamic and kinetic averages in computer simulations.

Such order parameters have been used recently in simulations of SPC/E water by Errington and Debenedetti [19]. The study found a number of interesting patterns in the relationship between structural order and water's anomalies. Distributions of the orientational order parameter at fixed density were found to be bimodal; as the temperature decreased, the high-tetrahedrality peak increased at the expense of the low-tetrahedrality one. Additionally, the average order parameter values exhibited a pronounced density dependence, leading to sharp maxima in orientational order and sharp minima in translational order upon compression. Furthermore, the two order parameters were found to be strictly correlated in the entire region of temperature and density where thermodynamic and transport anomalies occur, and where compression leads to a simultaneous decrease in translational and orientational order (the structurally anomalous region.) By strict correlation we mean the impossibility of changing the thermodynamic average of one order parameter without changing the other. This implies a mapping of the structurally anomalous region in the temperature-density plane onto a line in order parameter space. Furthermore, this line was found to be a boundary beyond which no state points could be found. In other words, the strict correlation between the order parameters in the structurally anomalous region is a locus of maximum tetrahedrality for any given amount of translational order.

In [19], an important relationship was found when the loci of structural extrema (maximum tetrahedrality, minimum translational order) were examined alongside kinetic (diffusivity) and thermodynamic (density) extrema in the temperature-density plane. The structurally anomalous region was found to enclose the region of kinetic anomalies (increase in diffusivity upon compression.) The boundary of thermodynamic anomalies (locus of density maxima) was further found to be contained inside the kinetically anomalous region. The resulting "cascade of anomalies" provides insight into the influence of order on anomalous behavior; it implies that structural order responsible for anomalies can be quantified and related to bulk thermodynamic and kinetic phenomena. In particular, because orientational and translational order were found not to be independent in the structurally anomalous region, either one could be used there as a single measure of structural order. Errington and Debenedetti introduced "isotaxis" lines, curves within the structurally anomalous region corresponding to a constant value of orientational, and hence, translational order. Locating the isotaxis lines tangent to the boundaries of thermodynamic and kinetic anomalies, they were able to determine the minimum amount of structural order necessary in order for these anomalies to occur.

The purpose of the present study is to extend the order parameter characterization of network-forming fluids to silica. As already mentioned, silica shares many of water's unusual thermophysical properties, in particular density and diffusivity maxima. By applying the Errington-Debenedetti approach to silica, the present study aims to explore the extent to which structural order parameters are general and accurate indicators of anomalous behavior in network-forming fluids. An important distinction to bear in mind, however, is the difference in intermolecular forces that promote formation of the network; these are hydrogen bonds in water, while in silica, the silicon-oxygen bonds are of a much stronger, covalent nature. Water molecules therefore remain distinguishable from each other in ice; however, silica forms network-spanning structures in its crystalline form. This difference suggests that results of structural analyses of silica and water should not necessarily be similar.

This paper is organized as follows. Section II reviews the order parameters and provides a summary of the computer simulation methods used in this work. Section III presents the results of the current investigation of silica, noting important comparisons with water. Finally, Section IV examines critically the generality of the order parameter approach for investigating liquid-phase anomalies in network-formers and suggests directions for future inquiry.

## II. Methods

*A. Characterization of molecular structural order*

Computer simulations easily lend themselves to detailed microscopic structural analysis. Consequently, local order in simulated network-forming fluids can be quantified by geometric means given a configuration of constituent particles. Both translational and orientational intermolecular correlations are in principle relevant measures of this order. The translational order parameter measures the tendency of pairs of molecules to be separated by preferential distances and is given by

$$t = \frac{1}{\xi_c} \int_0^{\xi_c} |g(\xi) - 1| d\xi \qquad (2.1)$$

where $g$ is the pair correlation function, $\xi = r\rho^{1/3}$ is a dimensionless distance expressed in units of the mean intermolecular separation, $\xi_c$ is a cutoff distance, and $\rho$ is the number density [20]. The translational order parameter vanishes for an ideal gas ($g = 1$), is large for crystal structures, and remains finite at the critical point. It is important to note that $t$ as defined here has no upper bound, and therefore is not normalized such that a value of 1 indicates perfect (i.e., crystalline) translational order. Alternative definitions of translational order are available which can be normalized and which do not require the pair correlation function. These alternatives may be constructed given knowledge of the preferred crystal structure and hence the occupation shells. We use equation (2.1) because it is a general form irrespective of the crystalline state and provides the same qualitative trends as crystal-specific definitions [20, 21].

The orientational order parameter measures the tendency of first-shell neighbor molecules to adopt a tetrahedral configuration about a central molecule. A convenient measure of orientational order in the vicinity of molecule $i$ is given by

$$q_i = 1 - \frac{3}{8} \sum_{j>k} \left[ \cos\theta_{ijk} + \frac{1}{3} \right]^2 \qquad (2.2)$$

where the summation is over the six possible pairs among molecule $i$'s nearest neighbors and $\theta_{ijk}$ is the angle formed between neighbors $j$ and $k$ and the central atom $i$ [22]. Contrary to $t$ in equation (2.1), $q$ is defined for each molecular center and therefore a distribution of $q$ values will be observed. It can be seen that $q$ is 1 when all angles are tetrahedral (i.e., $\cos(\theta_{ijk}) = -1/3$). When the positions of centers of molecules are uncorrelated, on the other hand, the mean value of $q$ is zero [19]. This limit corresponds to the ideal gas.

There is an important distinction between the use of the tetrahedral order parameter in silica and water. In both, we measure $q$ between molecular centers, involving silicon-silicon and oxygen-oxygen relationships, respectively. However, neither of the common stable forms of silica, $\alpha$- and $\beta$-quartz, possess strict silicon-silicon tetrahedrality [23]. Rather, each silicon and its four nearest oxygen neighbors are tetrahedrally arranged. In contrast, ice exhibits tetrahedral geometry between molecular centers, i.e., oxygen atoms. This would seem to raise questions about the use of a tetrahedrality measure for silica based only on the relative position of silicon atoms. It has been shown with the potential used in this work that the majority of silicon atoms are four-coordinated with their nearest silicon neighbors for T < 5000 K [6, 24]. Though the geometry may not be strictly tetrahedral, we obtain important trends on the evolution of structural order and its relationship to thermophysical properties by using the tetrahedrality parameter defined in equation (2.2), applied to Si-Si spatial correlations. We have also studied the behavior of $q$ defined between Si centers and O neighbors and have found that this $q$ gives rise to qualitatively similar results (see Results.)

The order parameters $t$ and $q$ are easily calculated during molecular simulation. Radial distribution functions are routinely tabulated, and from these $t$ is found. The calculation of $q$ requires one to locate the nearest neighbors of each molecule. Once neighbors are identified, $q$ is calculated for each molecular center at periodic intervals during the simulation. A histogram of $q$ values can then be calculated at the end of the run.

*B. Computer simulations of liquid silica*

To investigate the anomalous properties of liquid silica with order parameter mapping, extensive molecular dynamics simulations were performed similar to those previously done for water [19]. Two potentials for silica are commonly used in simulations, those developed by Tsuneyuki, Tsukada, Aoki, and Matsui (TTAM) [25] and by Van Beest, Kramer, and Van Stanten (BKS) [26, 27]. Both potentials have the same analytical form, with different fitted parameters. The TTAM potential was parametrized entirely from ab-initio calculations on small clusters of $SiO_4^{(4-)}$ and has been successful in reproducing polymorph behavior [28], diffusivity anomalies [10], and liquid and gas coordination structure [29, 30]. The BKS potential was parametrized from a combination of ab-initio data on $H_4SiO_4$ clusters and bulk crystal structure, and has been used extensively in previous liquid- and glassy-state studies and in relation to the energy landscape formalism [6-8, 31-35]. The BKS potential was chosen in this work so as to allow comparisons with these latter studies.

The BKS potential treats silica as independent silicon and oxygen ions, and has the form of an exponential-6 plus partial charge interaction between each of the three types of pairs. The form of the potential is

$$\phi_{ij}(r) = \frac{q_i q_j}{4\pi\varepsilon_0 r} + A_{ij} e^{-b_{ij} r} - \frac{C_{ij}}{r^6} \quad (2.3)$$

where $r$ is the ion-ion separation distance, $\varepsilon_0$ is the permittivity of free space, $q$ is a partial charge, and $A$, $b$, and $C$ are the parameterized constants. $A$ and $C$ are zero for silicon-silicon interactions, though the partial charges of $q_{Si} = 2.4e$ and $q_O = -1.2e$ provide Coulombic forces between all pairs. Because the potential exhibits negative divergence at small separation distances, a Lennard-Jones 30-6 correction is added with parameters optimized to minimize effects at large distances while preventing an inflection at short distances [7]. The corrected potential has the form

$$\phi_{ij}(r) = \frac{q_i q_j}{4\pi\varepsilon_0 r} + A_{ij} e^{-b_{ij} r} - \frac{C_{ij}}{r^6} + 4\varepsilon_{ij}\left[\left(\frac{\sigma_{ij}}{r}\right)^{30} - \left(\frac{\sigma_{ij}}{r}\right)^6\right] \tag{2.4}$$

where $\varepsilon_{ij}$ and $\sigma_{ij}$ are the energy and length parameters of the Lennard-Jones correction.

In comparing silica with water, one should bear in mind qualitative differences between the BKS potential for silica and the SPC/E potential for water [36]. The former deals exclusively with spherically symmetric, partially charged ions; the latter uses rigid intramolecular bonds and neutral molecules with an internal charge distribution. Therefore, it is impossible to unambiguously distinguish individual molecules in BKS silica, while individual water molecules are always identifiable in SPC/E water.

Molecular dynamics simulations were conducted in the NVE ensemble. 150 silicon and 300 oxygen atoms were placed in a cubic box under periodic boundary conditions. It has been claimed in previous studies of this potential that finite-size effects can be significant [37]; however, it has been shown that reliable diffusivities can be obtained provided the simulation is long enough [10]. We performed several test runs with varying system size, and these confirmed that the system size used in this work is sufficient to yield size-independent results. In particular, averages of diffusivity and order parameters for $N \geq 450$ did not change beyond numerical uncer-tainty. Although the pressure showed a slight N-dependence, the location of isochoric pressure minima (i.e., density maxima) showed no size dependence for N ≥ 450.

To properly treat the long-range electrostatic interactions, the Ewald summation was employed [38]. The $\alpha L$ parameter, which controls the rate of decay of the real-space potential, was set to 5.6. All $k$-vectors with magnitude less than or equal to $5 \times 2\pi/L$ were included in the reciprocal space summation, which was sufficient for energy conservation to one part in $10^5$. The non-Coulombic part of the potential was truncated and shifted at 5.5 Å. The masses of the silicon and oxygen ions were taken as 28.086 and 15.999 u, respectively.

Isochoric simulation runs were conducted at densities ranging from 1800 – 4200 kg/m$^3$. To improve statistics, results at each state point were averaged from three simulations at high densities (> 3000 kg/m$^3$) and from six simulations at low densities (≤ 3000 kg/m$^3$). The extra simulations at low density were necessary due to the onset of sluggish dynamics. Temperatures investigated were in the range 2500 – 6000 K. Periodic velocity rescaling was used during equilibration phases to bring the system to the specified temperature. A time step of 1 fs was used with the velocity-Verlet integration algorithm to solve for the atomic trajectories [38]. At a specified density, an initial configuration was generated by randomly placing all ions in the simulation box under the constraint that the force between any two pairs be below a specified cutoff. The initial equilibration then consisted of frequent velocity rescaling at 6000K for approximately 3000 time steps; this was necessary to overcome any high-energy configurations in the initial state. An equilibration phase then ensued for 25000 additional steps, with velocity rescaling every 50 steps. This was adequate for the system to lose memory of its initial configuration and equilibrate at the setpoint temperature. Subsequently, a production run of 200000 steps was used to gather the relevant data. The equilibration-production sequence was repeated at decrements of 250K for temperatures down to 2500K.

In addition to standard thermodynamic averages (temperature, pressure, energies), data collected during each production run yielded self-diffusion coefficients, radial distribution functions (RDFs), average translational ($t$) and orientational ($q$) order parameter values, and orientational order parameter distributions. Self-diffusivity was calculated using the Einstein relation and 100 time origins separated by 0.1 ps for averaging the mean-squared displacement with time. Though the diffusivity for silicon was consistently lower than that of oxygen, no difference in trends was observed; therefore, only oxygen diffusivities are reported. All structural parameters were averaged from values taken every 20 time steps during the simulation run, and are reported for silicon-silicon interactions only. The RDFs and $q$ distributions were created from histograms of 800 bins. Average $t$ values were calculated using the RDFs, and both averages and variances of $q$ values were calculated from the $q$ distributions.

For comparison with the NVE molecular dynamics results at low densities, we also conducted Monte-Carlo (MC) simulations in the NVT ensemble. In order to improve computational efficiency and speed, the MC simulations were performed such that the BKS ions were constrained to a cubic lattice [39]. A rather small lattice spacing of 0.05 Å was used to accommodate the implicit directionality of the interactions. For each state point, thermodynamic averages were taken from ten to fifteen runs of $15 \times 10^6$ particle move attempts. Each run was initially equilibrated at high temperature from a random atomic configuration, and similar to the MD simulations, a repeating equilibration-production procedure at fixed density was used to sample temperatures in decremental fashion.

## III. Results and discussion

Results for pressure calculation from simulations are reported in Fig. 1 for high densities and Fig. 2 for low densities. A minimum in the P vs. T curves, indicative of an isobaric density maximum, appears at 3200 kg/m$^3$ and persists as the density is lowered. However, considerable noise is present in the data at low densities due to the onset of slow dynamics and possible glass formation. To explore the loss of ergodicity and improve statistics, MD results were compared to those of the NVT Monte Carlo simulations. These latter results are shown alongside the molecular dynamics data in Fig. 2. It is clear that for the lowest densities, the system falls out of equilibrium since disparities between the two simulation methods become large. Furthermore, the dynamics are extremely slow in this region, and even the MC simulations do not adequately cover phase space in a reasonable number of steps, as shown by the large uncertainties.

Examining the MD/MC discrepancy, a cutoff temperature can be identified at each low density, below which calculated molecular dynamics averages are unreliable: 2750, 3000, 3000, 3250 K for 2800, 2600, 2400, and 2200 kg/m$^3$, respectively. The difference at densities below 2200 kg/m$^3$ is large for each temperature investigated, making suspect any data at those densities. Furthermore, it appears from both simulation methodologies that a limit of stability is reached around 2000 kg/m$^3$. That is, $(\partial P/\partial \rho)_T$ becomes negative at lower densities (data not shown.) This analysis shows that the determination of physical and structural properties at low enough densities and temperatures becomes very difficult due to the combined effects of slow dynamics and loss of stability. Although the anomalous behavior of liquid water was investigated near the liquid spinodal [19], silica's pronounced loss of ergodicity in the low-density, low-temperature region is a major difference between these two network-forming systems.

Our results for the diffusivity confirm the hypothesized glass formation and nonergodicity found in the pressure averages. Fig. 3 shows the time dependence of the mean-squared displacement for oxygen ions at several densities and temperatures. At 2000 kg/m$^3$, there is a clear onset of glasslike dynamics by 3250K. At the intermediate density of 3000 kg/m$^3$, the dynamics become faster due to the disruption of structural order in the system, but by 4000 kg/m$^3$ diffusion again becomes slow as a result of the high compression. The plot of diffusivity isotherms in Fig. 4 illustrates this trend. There is a diffusivity maximum at each temperature investigated. For T ≥ 4000 K, it is possible to detect a diffusivity minimum at lower density. However at low temperatures, this becomes difficult due to slow dynamics.

The distributions of orientational order, $q$, over the state points investigated reveal bimodal behavior similar to that reported for water [19] and are shown in Fig. 5. The results suggest that silicon atoms populate two distinct environments, one with high silicon-silicon tetrahedral order ($q\approx0.9$) and one with relatively low order ($q\approx0.45$). The population of tetrahedrally ordered silicon atoms increases as the temperature and density decrease. Similar effects were found in water [19]. The high-tetrahedrality peak disappears when the density increases to 4000 kg/m$^3$. At this point, temperature does not greatly affect the shape of the distribution because the system is completely constrained to one population.

A simple quantitative measure of the bimodality in $q$ is the variance in the distribution around the mean. Fig. 6 depicts the $q$ variance as a function of density for different temperatures. At low enough temperatures (T < 4750 K), there is a range of densities where compression leads to an increase in the variance of $q$. This reflects the gradual evolution of a unimodal, high-$q$ distribution into a bimodal distribution with high- and low-$q$ populations. Outside this intermediate density range, compression leads to a decrease in the variance of the tetrahedrality distribution. At low density, this reflects evolution towards optimal tetrahedrality in the approximate range $2100 < \rho < 2300$ kg/m$^3$. At high density the distribution becomes progressively sharper about the low-$q$ peak. At high enough temperatures (T > 4750 K), on the other hand, compression leads only to a narrowing of the tetrahedrality distribution around the low-$q$ peak. Interestingly, it appears that the maxima in the variance of $q$ occur near isobaric density maxima (results not shown.) Further work is being pursued to clarify this unexpected relationship.

The order parameter map generated from isotherms of $q$ vs. $t$ is shown in Fig. 7. Similar to water [19], upon compression the system experiences a maximum in $q$ followed by a minimum in $t$. Additionally, like water, the silica order parameter map suggests an inaccessible region in the high $q$, low $t$ quadrant. That is, for any given value of $t$, there is a maximum value of mean tetrahedrality that the system can attain. Likewise, for any given value of mean tetrahedrality, there is a minimum attainable translational order. There is, however, a qualitative difference in the shape of the $t$ minimum with respect to that seen in water. Whereas in water this minimum is sharp, silica's is noticeably broad. Furthermore, the portions of each isotherm between maximum tetrahedrality and minimum translational order do not collapse onto a single line, as is the case for water. Thus in the anomalous region where both translational and orientational order decrease upon compression, $t$ and $q$ are not perfectly correlated. This degree of independence renders impossible an "isotaxis" characterization such as was found useful in water [19]. It is important to note differences in the potentials used in water and silica in this context. One might hypothesize that the rigid structure of SPC/E molecules affects the nature of the $t$ minima, and that a flexible bond model (as is implicit in the BKS potential) might produce a broader extreme. This possibility is under further investigation.

It may also be important that the silicon-silicon relationships in crystalline silica (quartz) are not strictly tetrahedral [23]. That is, $q$ in the ground-state network is not 1, but slightly less. We therefore also investigated $q$ between silicon centers and oxygen neighbors, which has a value very near 1 in the crystalline state [23]. The two $q$'s exhibited qualitatively similar trends and the isothermal maxima in $q_{Si-O}$ occurred at the same density as the corresponding $q_{Si-Si}$ within numerical error (results not shown.)    There is some ambiguity about the underlying structural phenomena that $q_{Si-O}$ measures.   The electrostatic repulsion between oxygen atoms immediately surrounding a silicon center strongly favors a tetrahedral arrangement.  We find that near its maximum, $q_{Si-O}$ varies significantly less with density than $q_{Si-Si}$.  This makes determination of the $q$ maximum in the former more difficult and shows that it is less sensitive to local order than the latter.  Therefore, it appears that if a single orientational order parameter is to be used, the appropriate $q$ should measure "local" relationships (Si-Si), but not immediate relationships (Si-O).

The loci of density, diffusion, and order parameter extrema are shown in Fig. 8. Due to glasslike behavior at low temperatures and densities, data for density maxima at very low densities, and $q$ maxima and diffusivity minima at very low temperatures are not included. This does not preclude analysis of the main trends. In water, a cascade of anomalies was observed in which the region of structural anomalies (where $q$ and $t$ decrease upon compression) completely contained the region of kinetic anomalies (where diffusivity increases upon compression), which in turn enclosed the region of thermodynamic anomalies (where density decreases upon isobaric cooling.)   In silica, there is an inversion of the structurally and kinetically anomalous regions. That is, transport anomalies are not preceded by anomalies in microscopic structural order. Despite this difference, the qualitative relationship between kinetic and thermodynamic anomalies in water and silica is the same. This similarity is particularly notable because of the significant differences in the nature of intermolecular bonding in these fluids. The appearance of diffusivity anomalies before density anomalies may therefore be a common feature of fluids that form tetrahedral networks, regardless of the mechanism by which they do so.

Insofar as the order parameters used in this work do not anticipate the occurrence of diffusivity anomalies (as was the case for water [19]), we conclude that they do not provide complete information about the microscopic causes of anomalous bulk behavior in silica. One might expect this to be the case for the orientational order parameter since, as mentioned, the two forms of quartz do not exhibit strict silicon-silicon tetrahedrality. However, the relationship between translational order minima and diffusivity maxima is also inverted in the $(T, \rho)$ plane—the latter precedes the former. This suggests a cause common to both order parameters. There is some flexibility in the selection of order parameters for the examination of structural relationships; average $q$ and $t$ values can be examined for all pairs of atom types. Therefore, the possibility exists that the most appropriate measure of local order is obtained not from one, but a combination of the parameters for each pair type. Whether or not this is actually the case remains a subject for further investigation.

## IV. Conclusion

The similarities in the relationship between order parameters and anomalous behavior in water and silica are a consequence of the network-forming character of both fluids. As was previously found to be the case for water, the current simulation results for silica reveal a bimodal distribution of orientational order; diffusivity and density extrema; minima in translational order; and maxima in orientational order. Differences between the two fluids arise when comparing the translational order minima (silica's are much broader) and the regions of anomalous behavior (silica's kinetic anomalies are not preceded by structural ones in the temperature-density plane.) Further, there may be a connection between silica's orientational order bimodality and its density anomalies. The equivalent trend in water has not been studied. Silica's distinctions might be attributed to its atomic (rather than molecular) potential, non-tetrahedral silicon-silicon crystalline arrangement, and extreme viscous slow-down at low temperatures.

A natural extension of this work is related to the energy landscape formalism of dynamic systems [40]. Here, the configurational entropy measures the number of distinct potential energy minima a system samples, between which structural rearrangement is required. It is logical to suggest that configurational entropy is related to microscopic structural order in network-forming fluids. These substances' highly directional intermolecular interactions will have a direct impact on the topography of their potential energy landscapes (the multidimensional energy hypersurface formed by a system's structural degrees of freedom.) The signature of these interactions will manifest itself in the behavior of the configurational entropy, which in contrast to simple liquids, should delimit an anomalous region in state space defined by landscape statistics.

Scala and co-workers investigated SPC/E water in this context [41]. Through extensive molecular dynamics calculations, they found a correspondence between diffusivity maxima and configurational entropy maxima. At temperatures and densities where compression enhanced diffusion, the same applied to configurational entropy. Though diffusivity minima were not investigated, their results provided substantial support for a connection between the statistical properties of the energy landscape and kinetic anomalies. In the same spirit, Saika-Voivod and co-workers calculated the configurational entropy for BKS silica at several densities and showed a relationship between inflections in configurational entropy vs. temperature, maxima in isochoric specific heat, and maxima in isothermal compressibility [8]. An important finding in this work was identification of landscape signatures of a fragile-to-strong transition, as evidenced by the temperature dependence of the configurational entropy.

These studies encourage future investigation of the connection between molecular structural order and the energy landscape. Knowledge of the behavior of configurational entropy over a greater range of densities may permit mapping of the landscape-influenced anomalous region in silica (where increases in density cause increases in configurational entropy.) Given the close connection found for water, configurational entropy may prove to be a more general "order parameter" than molecular structural ones. Furthermore, future studies might consider the behavior of structural order parameters in the vicinity of a fragile-to-strong transition like that observed in silica. The results would clarify the causes and generality of such a transition.

## V. Acknowledgements

The authors would like to thank Jeff Errington for helpful discussions. MSS is supported by a Hertz Foundation Graduate Fellowship. We gratefully acknowledge the support of the Department of Energy, Division of Chemical Sciences, Geosciences, and Biosciences, Office of Basic Energy Science (grants DE-FG02-87ER13714 to PGD and DE-FG02-01ER15121 to AZP.)

**Figures**

FIG. 1. Temperature dependence of pressure from molecular dynamics simulations at high densities. A temperature of minimum pressure appears for $\rho \leq 3200$ kg/m$^3$, corresponding to a constant-pressure density maximum anomaly. A typical error bar is shown.

FIG. 2. Temperature dependence of pressure at low densities. Lines are from molecular dynamics and symbols are from Monte Carlo NVT simulations. Good agreement is obtained at high temperatures and densities; however, the results disagree at low temperatures and densities, indicating the nonergodic behavior. In both cases instability occurs below 2000 kg/m$^3$ where $(\partial P/\partial \rho)_T > 0$. Molecular dynamics error bars have been omitted for clarity.

FIG. 3. Mean-squared displacement of oxygen atoms. At low density, network formation limits diffusion and gives rise to glasslike dynamics at low temperatures. An intermediate density regime of enhanced diffusion occurs due to disruption of order, but as the density continues to increase, diffusion again decreases.

FIG. 4. Self-diffusion coefficient for oxygen ions as a function of density. A clear maximum exists at every temperature. The minima are less clear at low temperatures due to their proximity to the unstable and slow-dynamics region. Several low-density data points have been omitted at low $T$ due to uncertainties arising from very slow dynamics.

FIG. 5. Frequency distribution for the orientational order parameter, $q$. Arrows indicate the direction of increasing temperature, from 2500 to 5500 K. The system exhibits a bimodal popula-

tion in which highly ordered configurations are populated at low density and low temperature, and less ordered configurations are populated at higher density.

FIG. 6. Isotherms of variance in $q$ calculated from the order parameter distribution. For temperatures below 4750K, a minimum and maximum in the variance result during compression as the population switches from the high-tetrahedrality peak to the low-tetrahedrality peak in the bimodal distribution.

FIG. 7. Isotherms of orientational order ($q$) versus translational order ($t$). Along each isotherm, density decreases monotonically from the low-$q$ extreme (left) in decrements of 200 kg/m$^3$ for $4000 \leq \rho \leq 2600$ kg/m$^3$ and of 100 kg/m$^3$ for $2600 \leq \rho \leq 1800$ kg/m$^3$. There exist a broad minimum in $t$ and a sharp maximum in $q$; the former contrasts sharply with water, for which there is a sharp $t$ minimum. Noise in the data at the top-right portion of the figure is due to glasslike dynamics at low densities and temperatures.

FIG. 8. Loci of anomalies for BKS silica. The TMD points define the locus of isobaric density maxima, inside of which the coefficient of thermal expansion is negative (thermodynamic anomalies.) $D_O$ define the locus within which the diffusivity of oxygen increases upon isothermal compression (kinetic anomalies.) Contrary to what was found for water, the structurally anomalous region (bounded by the locus of $q$ maxima and $t$ minima, within which both measures decrease upon isothermal compression) does not enclose the region of kinetic anomalies.

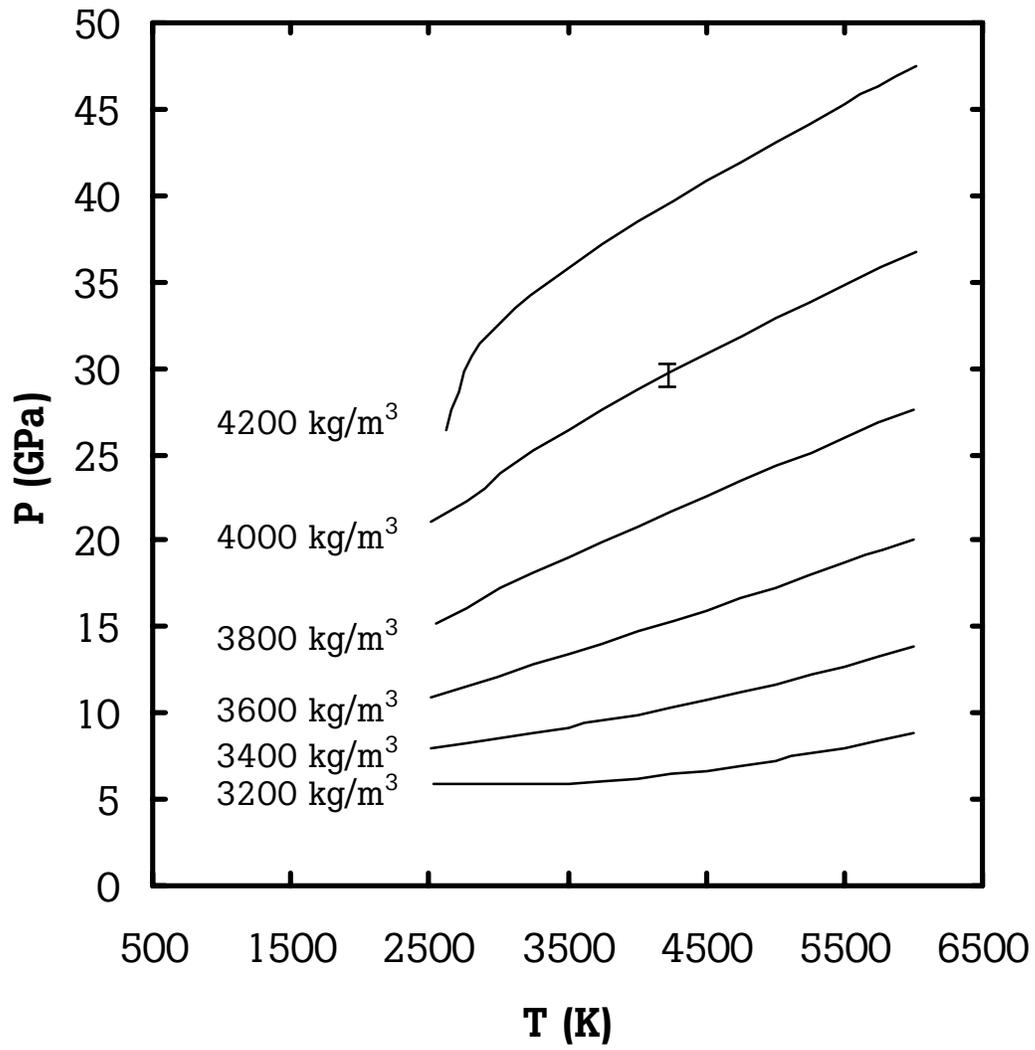

Figure 1

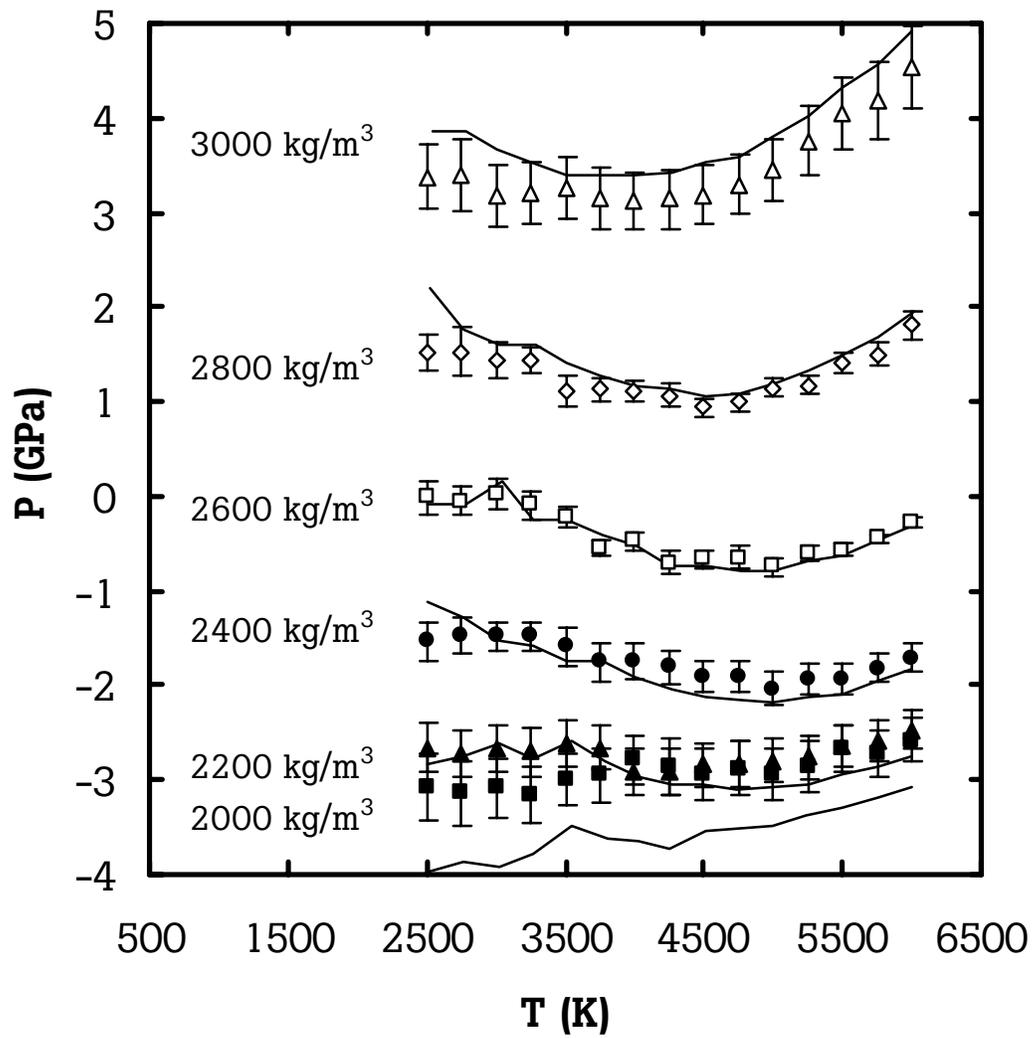

Figure 2

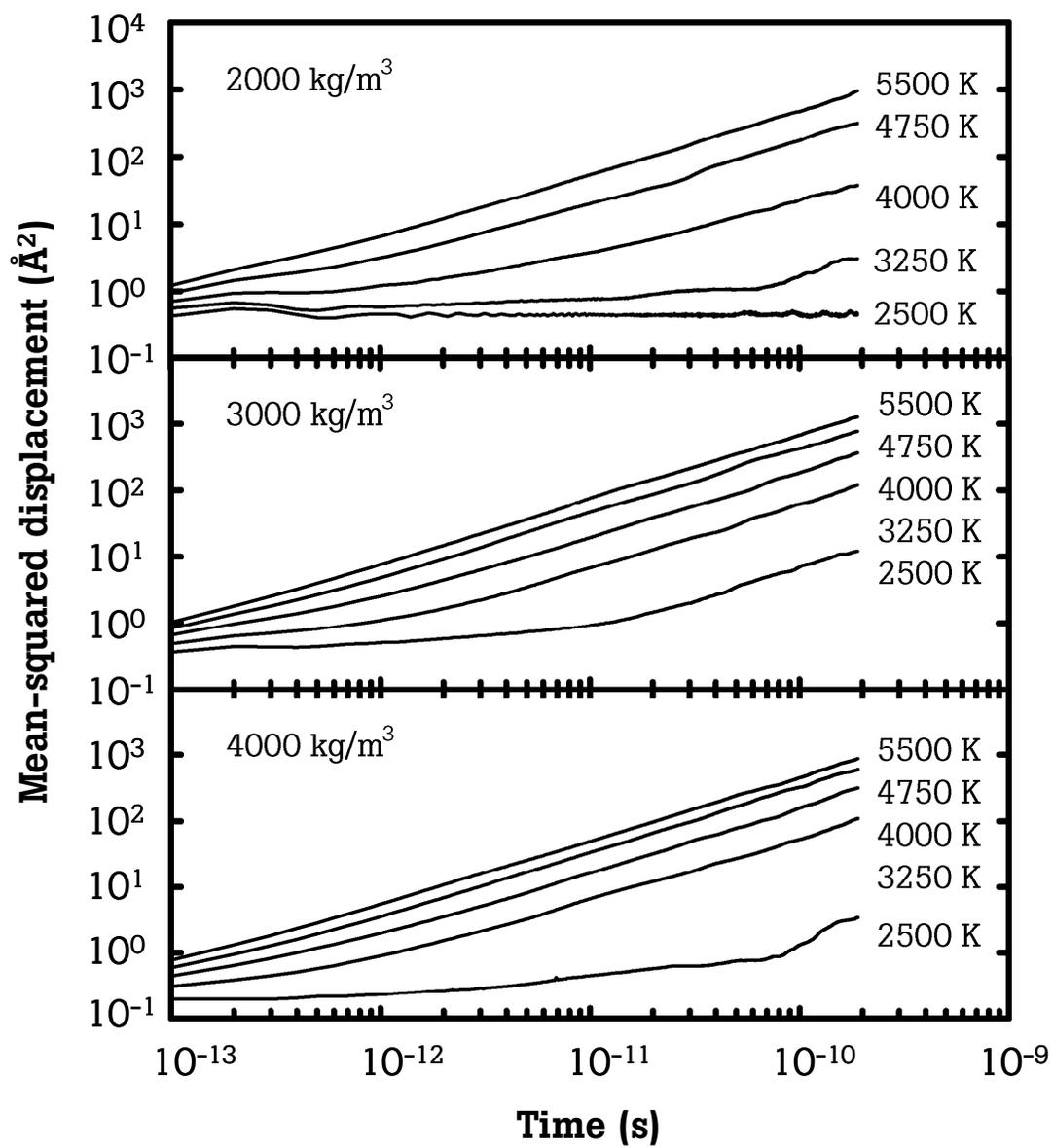

Figure 3

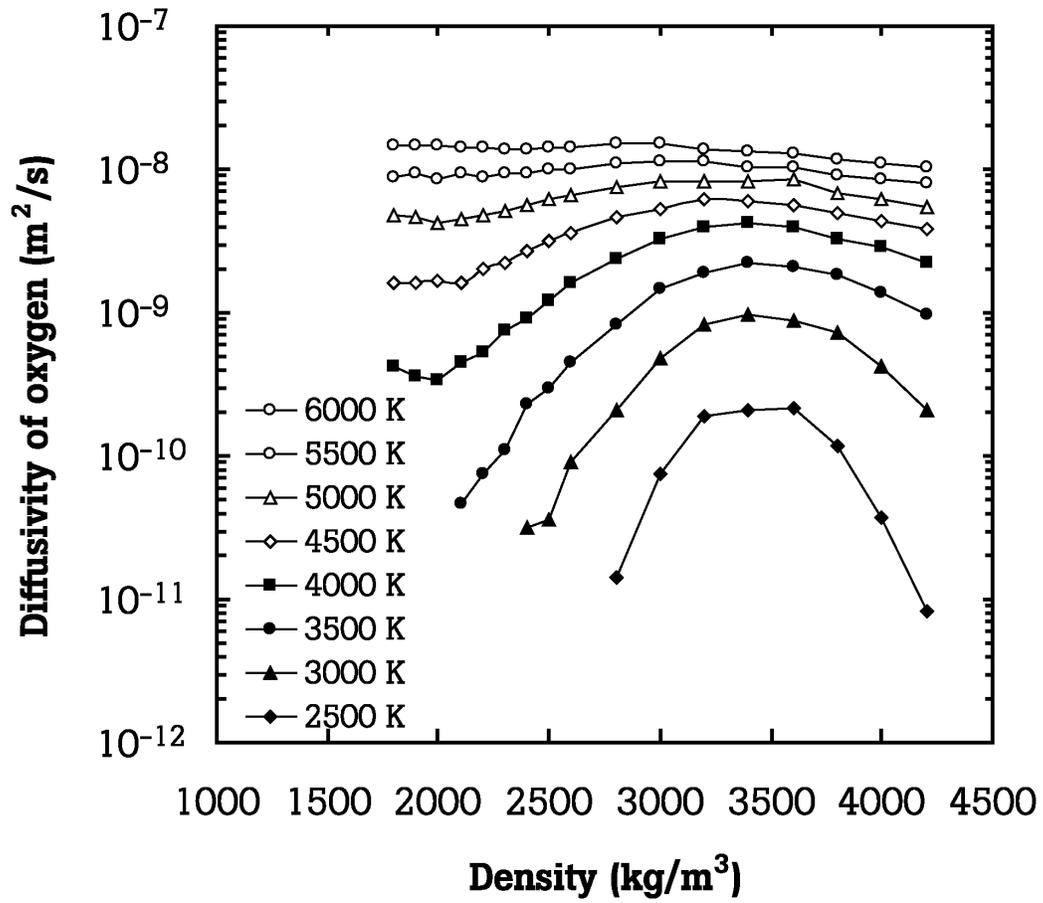

Figure 4

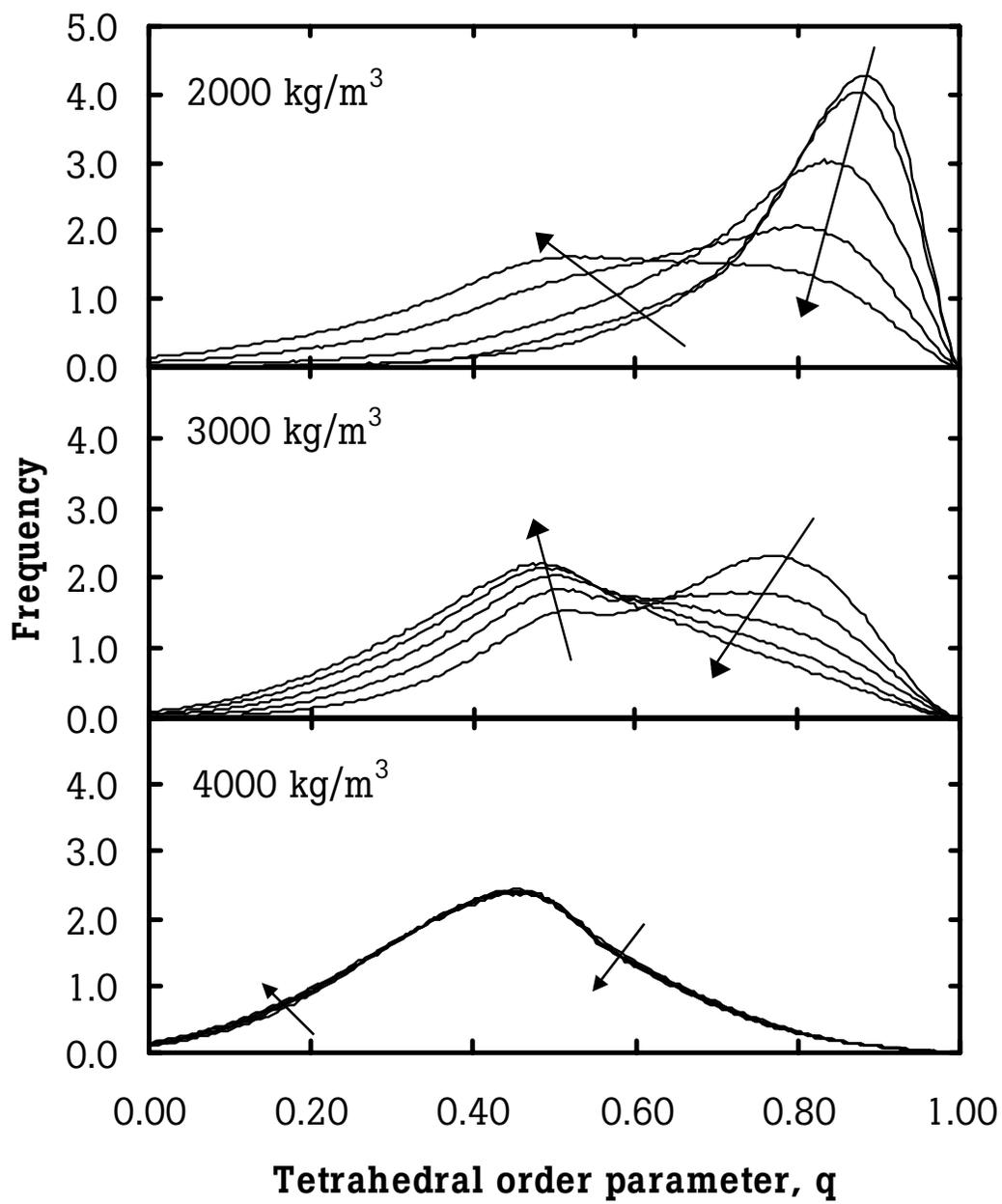

Figure 5

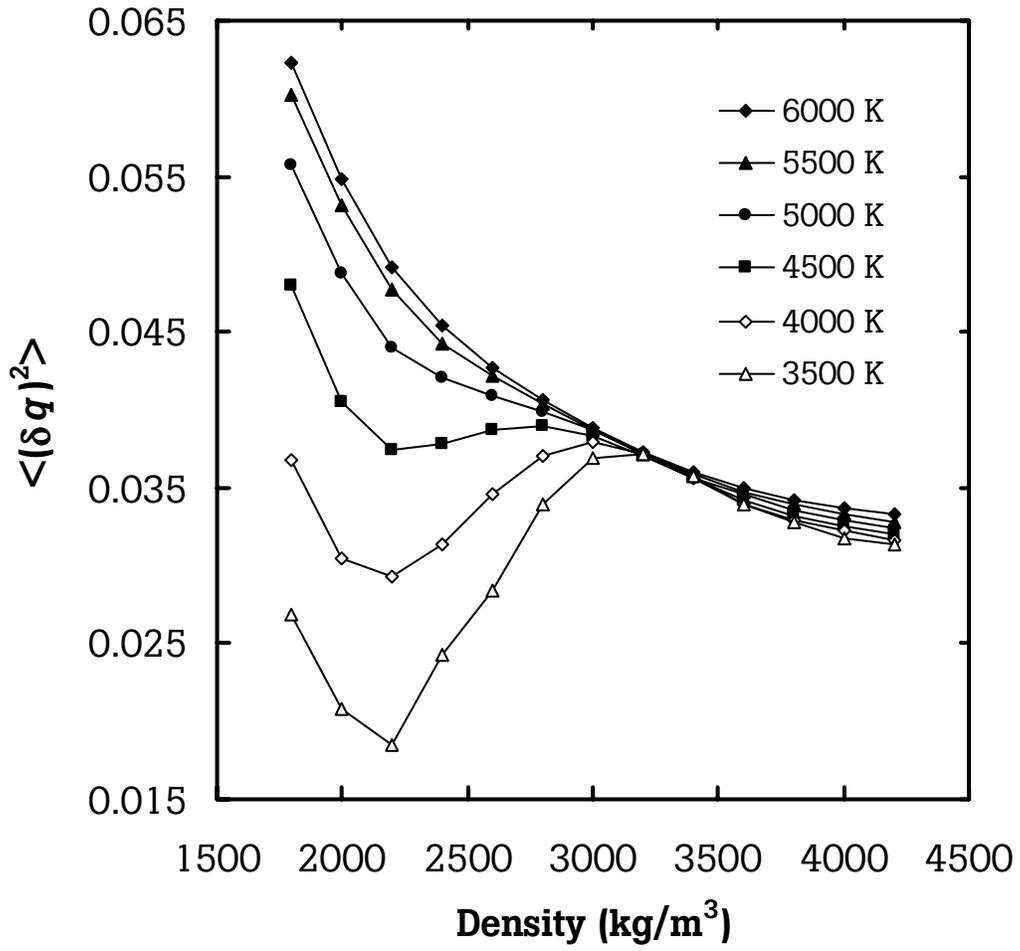

Figure 6

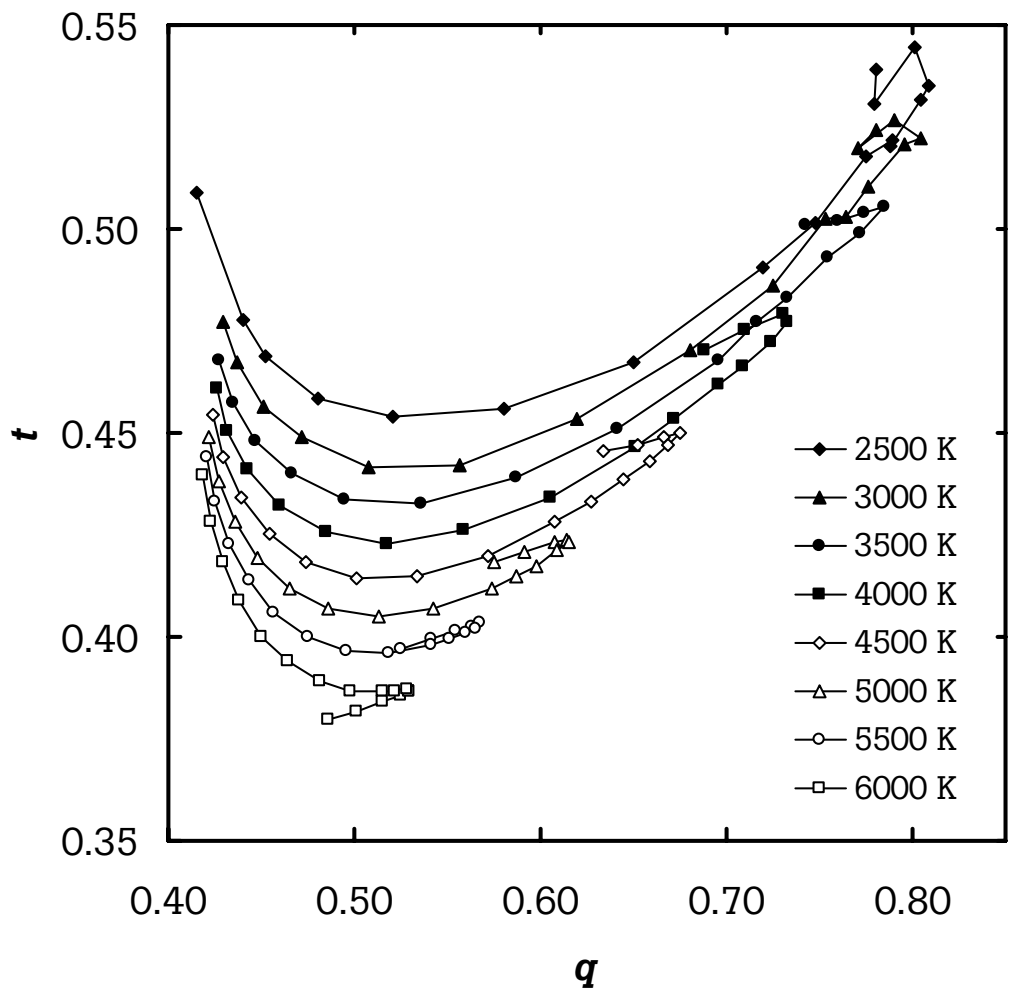

Figure 7

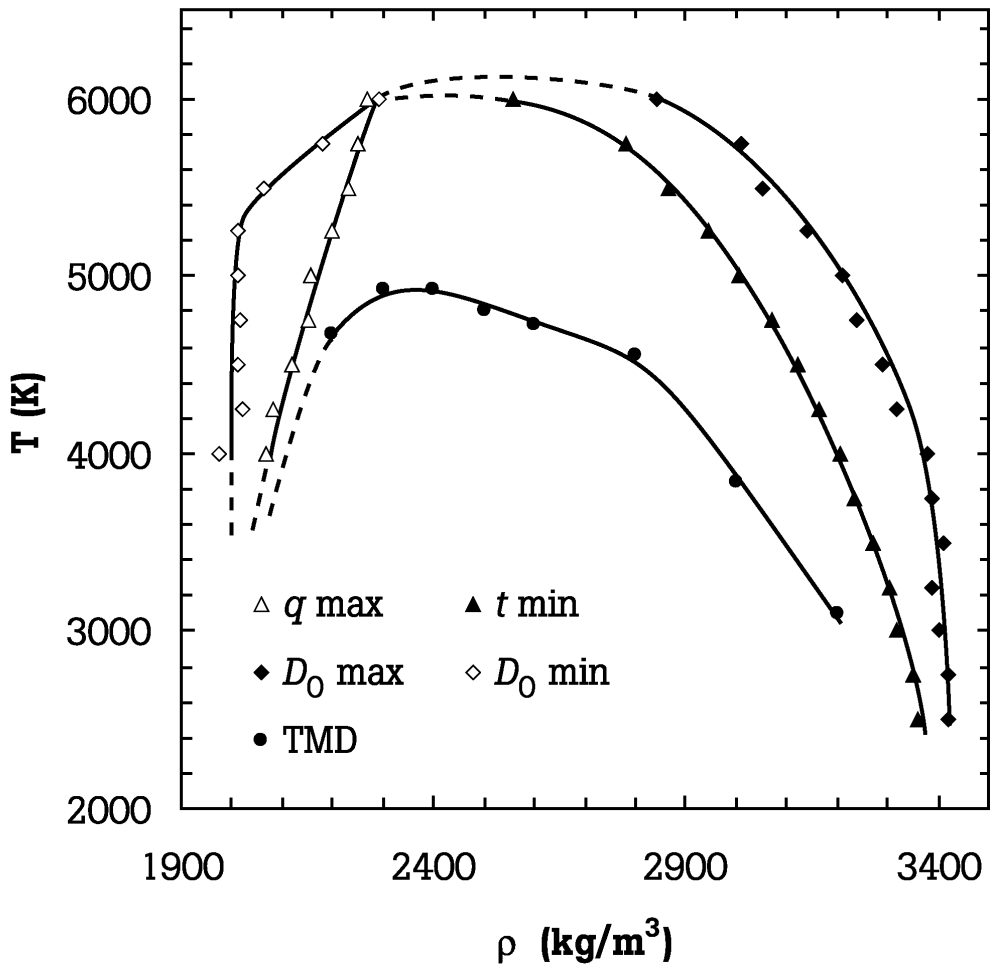

Figure 8